\documentclass[superscriptaddress, showpacs, preprintnumbers, aps, twocolumn, prb, floatfix]{revtex4-1}
\usepackage{graphicx}
\usepackage{dcolumn}
\usepackage{amsmath}
\usepackage{bm}

\begin{document}

\title{Dynamical Corrections to Spin Wave Excitations in Quantum Wells due to Coulomb Interactions and Magnetic Ions}
\author{Cynthia Aku-Leh\footnotemark{} }
\affiliation{Institut des Nanosciences de Paris, UMR 7588, CNRS/Universit\'e Paris VI et
VII, Campus Boucicaut, 140 rue de Lourmel, 75015 Paris, France}
\affiliation{Department of Physics, King's College London, Strand, London WC2R 2LS,
United Kingdom}
\author{Florent Perez}
\affiliation{Institut des Nanosciences de Paris, UMR 7588, CNRS/Universit\'e Paris VI et
VII, Campus Boucicaut, 140 rue de Lourmel, 75015 Paris, France}
\author{Bernard Jusserand}
\affiliation{Institut des Nanosciences de Paris, UMR 7588, CNRS/Universit\'e Paris VI et
VII, Campus Boucicaut, 140 rue de Lourmel, 75015 Paris, France}
\author{David Richards}
\affiliation{Department of Physics, King's College London, Strand, London WC2R 2LS,
United Kingdom}
\author{Grzegorz Karczewski}
\affiliation{Institute of Physics, Polish Academy of Sciences, al. Lotnik\'ow 32/46,
02-668 Warszawa, Poland}
\date{\today }

\begin{abstract}
We have measured dispersions of spin-flip waves and spin-flip
single-particle excitations of a spin polarized two-dimensional
electron gas in a Cd$_{1-x}$Mn$_x$Te quantum well using resonant
Raman scattering. We find the energy of the spin-flip wave to be
below the spin-flip single particle excitation continuum, a
contradiction to the theory of spin waves in diluted magnetic
semiconductors put forth in [Phys. Rev. B \textbf{70} 045205
(2004)]. We show that the inclusion of terms accounting for the
Coulomb interaction between carriers in the spin wave propagator
leads to an agreement with our experimental results. The dominant
Coulomb contribution leads to an overall red shift of the mixed
electron-Mn spin modes while the dynamical coupling between Mn ions
results in a small blue shift. We provide a simulated model system
which shows the reverse situation but at an extremely large magnetic
field.

\end{abstract}

\footnotetext[1]{Current affiliation: Max Born Institute, Max Born Strasse 2A, Berlin 12489.}
\maketitle

\section{\label{intro}Introduction}

Collective spin dynamics in dilute magnetic semiconductors (DMS) is an active field under intense investigations.\cite
{UltrafastGaMnAs09, GilbertFerro, Vladimirova(2008), PerakisPRL08} This field provides an insight into the origins of carrier-induced
ferromagnetism in semiconductors \cite{FerroDMSDietl, MacDonaldPRL2000, KoenigPRLEPR03, PerezPRL07,Vladimirova(2008),TeranPRL2003} and an understanding of particular features of the DMS \cite{Mauger83, MacDonaldPRL2000} due to the presence of two spin
sub-systems that are dynamically coupled via Coulomb-exchange interaction: that of the itinerant carrier and that of the
localized magnetic impurities. As an example of these features, the transverse spin excitation spectrum has been
theoretically found to be composed of three types of excitations. These are: two collective excitations corresponding
to itinerant and localized spins precessing in phase or out of phase to each other, and single-particle (or Stoner-like)
excitations of the itinerant carriers.\cite{Mauger83, MacDonaldPRL2000,FrustagliaPRB04} The in-phase collective mode is the
Goldstone-like mode. The out of phase mode has a dominant contribution from the itinerant carrier subsystem.\cite
{MacDonaldPRL2000, FrustagliaPRB04} Experimental evidences of the entire spectrum in ferromagnetic DMS like GaMnAs is not
available. Reported so far are features related to the zone-center in phase mode, dominated by the Mn spin
precession, its dynamics\cite{MagnonGaMnAs_Wang07, HashimotoPRL08,UltrafastGaMnAs09} and its ferromagnetic resonance.\cite
{GaMnAsFMR} We find no experimental data available for the out of phase mode. Indeed, ferromagnetism in GaMnAs systems
requires a high Mn concentration, which destroys the quality of the crystal potential and smooths out all optical resonances.

To gain more insight into the DMS spin excitation spectrum, Cd$_{1-x}$Mn$_x$Te doped quantum wells are a very good alternative as they
are clean and efficient to capture the general properties of the collective spin dynamics in DMS materials. The high quality
of Cd$_{1-x}$Mn$_x$Te quantum well structures leads to a high mobility of the itinerant gas (electrons or holes) together with well-defined optical resonances,\cite{BoukariPRB06} an important condition for the observation of electronic Raman signals
from the carrier subsystem. \cite{JusserandPRL03} The presence of Mn impurities interacting with the two dimensional electron
gas (2DEG) embedded in the Cd$_{1-x}$Mn$_x$Te quantum well, thus, provides interest on two levels. First, the static mean-field \textit
{s-d} contribution from the Mn ions produces a large Zeeman splitting:
\begin{equation}
Z_{e}\left(B\right) =\Delta -\left\vert g_{e}\right\vert{\mu}_{\text{B}}B,
\label{eqZe}
\end{equation}
where $\Delta$ is the Overhauser shift due to the Mn spin polarization and $-\left\vert g_{e}\right\vert{\mu}_{\text{B}}B$ is the normal 
Zeeman splitting of the conduction band (here $g_{e}\simeq-1.64$ for $\mu_{\text{B}}>0$). This leads to a high spin polarization degree $\zeta =\left( n_{\uparrow }-n_{\downarrow }\right)/n_{
\text{2D}}$ in the 2DEG.\cite{PerezPRB2009} It makes it a suitable system to study features of spin-resolved Coulomb
interaction.\cite{PerezPRL07, AkulehPRB07} The above approach was followed in our previous work to study spin excitations
of the spin-polarized 2DEG (SP2DEG).\cite{JusserandPRL03, PerezPRL07, PerezPRB2009, GomezPRB10}. Propagative spin flip waves
(SFW) of electrons were observed \cite{JusserandPRL03} and their appearance was linked to the Coulomb interaction between
electrons.\cite{PerezPRB2009} On a second level, it provides evidence of the mixing between electrons and Mn spin modes in the
frequency\cite{TeranPRL2003} and time\cite{Vladimirova(2008)} domains as predicted in Ref.\cite{KoenigPRLEPR03}. Nevertheless,
a connection between these manifestations needs to be made: in the first approach, \textit{s-d} dynamical coupling with the
Mn was neglected, while in the second, Coulomb interaction between electrons did not manifest experimentally and was not taken
into account in theories of Ref.\cite{KoenigPRLEPR03, FrustagliaPRB04}. There is, therefore, a lack of full understanding of the mixed
spin-excitations where both Coulomb interactions between electrons and the \textit{s-d} dynamical coupling with Mn spins
interplay. Moreover, in the later studies,\cite{TeranPRL2003, Vladimirova(2008)} only the zone center collective electron-Mn
modes were probed. From the Larmor's
theorem,\cite{JusserandPRL03} it is known that the spin system
behaves as if the electrons were not interacting through Coulomb.
This explains why theories describing the electron-Mn modes, without
the Coulomb interaction between carriers,\cite{KoenigPRLEPR03}
agree well with the experimental data of Ref.\cite{TeranPRL2003}.
Out of zone center, predictions of Ref.\cite{FrustagliaPRB04}, found
the out of phase mode (so-called stiff mode) to have a positive
dispersion above the Stoner continuum. However, we have shown by
angle-resolved Raman measurements that the collective electron spin
mode (SFW mode), was below the spin-flip single-particle excitations
(SF-SPE)\cite{PerezPRL07}, together with a negative slope dispersion
with the Raman transferred in plane momentum $q$.\cite
{JusserandPRL03} It is natural to make the correspondence between
the stiff mode of Ref.\cite{FrustagliaPRB04} and the SFW mode of
Refs.\cite{PerezPRL07} and \cite {JusserandPRL03} (Stoner-like
excitations are obviously the SF-SPE). Therefore, our experimental observations
are in contradiction with the theory of Ref.\cite{FrustagliaPRB04}.

In the following we will use the denomination "out of phase mode" or
OPM to describe a mode which coincides with the stiff mode when the
Coulomb interaction between carriers is neglected or to the SFW mode
when the \textit{s-d} dynamical coupling with Mn spins is neglected.

In this work, we want to resolve this contradiction, by first
showing new experimental data that clearly shows the SF-SPE
continuum and that the OPM propagates below it with a negative
dispersion. Secondly, we will introduce a new propagator for the
electron-Mn collective modes that includes the Coulomb interaction
between spin-polarized carriers. Finally, we justify the
experimental features by comparing the \textit{s-d} exchange
dynamical correction due to the coupling of the electron and Mn spin
precession with the correction introduced by the Coulomb interaction
between carriers. We will demonstrate that the ratio of these two
corrections does not depend, at standard conditions, on the Mn
concentration and that the Coulomb correction always dominates over
the \textit{s-d} dynamical shift except for very large electron
densities.

\section{\label{expt}Samples and experimental setup}

Our measurements were made in the back scattering geometry on a Cd$_{1-x}$Mn$_{x}$Te single quantum well with electron
densities $n_{\text{2D}}$ ranging from $2\times 10^{11}$ cm$^{-2}$ to $4\times 10^{11}$ cm$^{-2}$ and an effective Mn$^{2+}$
concentration $x_{\text{Mn}}$ ranging from $0.24\%$ to $1\%$.\cite{AkulehPRB07} The quantum wells were grown by molecular beam
epitaxy on a GaAs substrate.\cite{karczewski} These quantum wells and their spin polarizations have been characterized in our
previous publication.\cite{AkulehPRB07} Here we will consider samples A and B of that reference. Both A and B have a density of $n_{
\text{2D}}=2.9\times 10^{11}$ cm$^{-2}$ and different Mn concentrations, $0.46\%$ and $0.75\%$, respectively. Measurements were taken with
the samples immersed in superfluid helium at $\sim 1.5$ K inside a superconducting magnetic cryostat. The essential part of
the cryostat is in the rod holding the sample, which allows for control in three directions, including rotation of the sample
in the plane of the magnetic field for dispersions while the collection/reflection mirror was kept fixed. This design aids in
an easy study of the Raman scattering dispersions with an accuracy of 1 degree ($ q=0.3\times 10^{6}$ $\mu m^{-1}$). To
determine the in-plane wave vector $q$ transferred in the plane of the quantum well, the angle of rotation $\theta $
is measured between the normal of incidence of the sample and the scattered light. $q$ is obtained from the geometry of the
set-up as $q\sim \frac{4\pi }{\lambda }\sin {\theta }$, where $\lambda $ is the excitation wavelength.\cite{angle} The holder
was inserted in the core of a solenoid producing magnetic fields of up to 5 T in the plane of the quantum well. A tunable cw
Ti-sapphire laser, pumped by an Ar$^{+}$ laser was used as light source. The laser power density was kept below 0.1 W/cm$^{2}$
 and the linewidth was 0.3 cm$^{-1}$ at a slit opening of 50 $\mu $m. The energy of the laser was tuned to resonate close to
the transition between the first conduction band and the second heavy hole band of the quantum well.

\section{\label{RD} Results and discussion}

\begin{figure}[tbp]
\includegraphics[width=\linewidth]{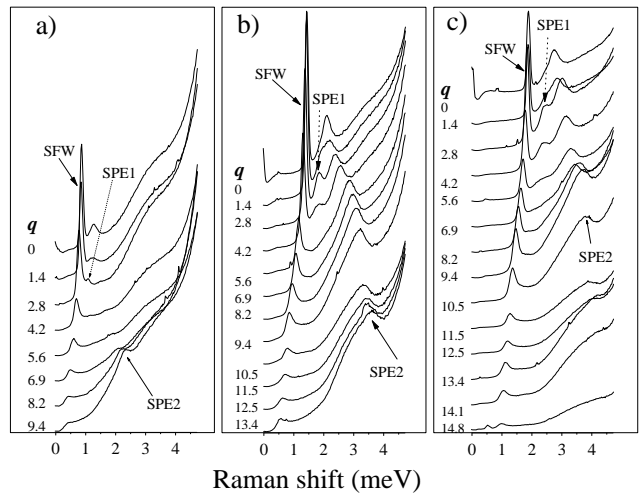}
\caption{Raman spectra of collective SFW and SF-SPE for magnetic fields: (a) $B=0.6$ T, (b) $B=1.2$ T and (c) $B=2.1$ T in sample A.
Arrows indicate locations of SF-SPE, SPE1 and SPE2 as described in the text. The lines at 0 meV are remnants of the
laser line.}
\label{fig:spectra}
\end{figure}

In Fig.~\ref{fig:spectra}, we show typical Raman scattering spectra obtained on sample A for various Raman transferred momentum $q$
obtained at different magnetic fields: (a) $B=0.6$ T, (b) $B=1.2$ T and (c) $B=2.1$ T. In the figure, $q$ varies from 0 to $\sim $15
$\mu $m $^{-1}$. All spectra are dominated by a sharp-peaked line associated with the collective SFW as already demonstrated in
Ref.\cite{JusserandPRL03}. We will show in the following that this mode is exactly the out of phase electron-Mn mode (the stiff mode of
Ref.\cite{FrustagliaPRB04}) of the DMS system. The SFW mode is narrow in width compared to the other excitations in the spectra and is
found to be intense at all magnetic fields. As $q$ is increased, the intensity and energy of the SFW decreases and eventually diminishes
at a critical wave vector, $q_{c}$.\cite{PerezPRB2009} The linewidth of the SFW broadens with increasing $q$, damping out due to
interactions with the SF-SPE continuum.\cite{GomezPRB10}

At high energies, features of the SF-SPE are observed in the spectra of Fig.~\ref{fig:spectra}. These features emerge as
the magnetic field is increased, showing a fan-shaped like behavior at large $q$ values leading to a continuum. This continuum
corresponds to the joint density of state of electron-hole pairs in
the Fermi sea which has a characteristic double-peak
structure.\cite{PerezPRB2009} In the following, we have labeled the
maxima positions of the SF-SPE continuum as SPE1 and SPE2. In the
long wavelength limit, SPE1 and SPE2 peaks are close to $Z^{\ast
}\pm\hbar v_{\text{F}\uparrow}q$ where $v_{\text{F}\uparrow}$ is the
Fermi velocity of the minority spin subband and $Z^{\ast }$ is the
renormalized Zeeman energy.\cite{PerezPRL07} SPE1 and SPE2 are
degenerate at $q=0$ forming a single peaked structure positioned at
$Z^{\ast }$.\cite{PerezPRL07} At small wavevectors, a clear
distinction between SPE1 and SPE2 is difficult as these features
overlap. SPE1 shifts to low energies, decreasing in intensity and
eventually vanishing as $q$ is increased. SPE2 on the other hand,
increases both in intensity and energy when increasing $q$. The dip
between SPE1 and SPE2 is due to the occupancy of the upper spin
subband.

\begin{figure}[tbp]
\includegraphics[width=\linewidth]{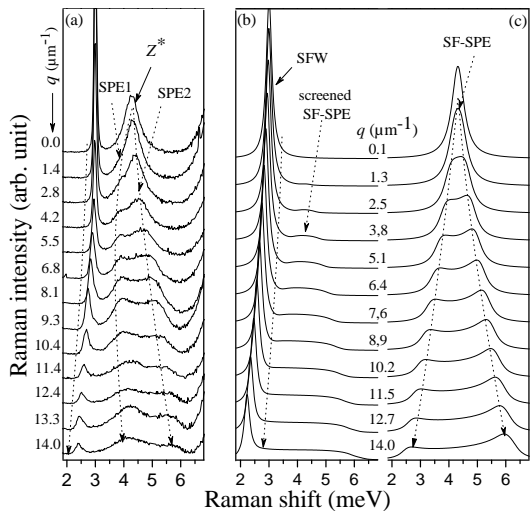}
\caption{Comparison between Raman spectra (a) at different $q$ values taken on sample B with the 
corresponding calculated spectra (b) and (c). (b) is obtained from the collective response and (c) 
from the single particle response.}
\label{fig:SampB&Theory}
\end{figure}

In Fig.~\ref{fig:SampB&Theory}, we compare the Raman scattering
spectra taken on sample B ($x_{Mn}=0.75\%$ and $n_{
\text{2D}}=2.9\times 10^{11}$ cm$^{-2}$) for various $q$ with
calculated spectra for the same conditions obtained from the
collective response and the single particle response of Ref.\cite
{PerezPRB2009}. The experimental spectra are neither given by the
collective response, nor the single-particle one, but by a mixing of
both responses. This is a consequence of the optical resonance with
the incoming Raman photon.\cite{SarmaPRL99}

\begin{figure}[tbp]
\includegraphics[width=\linewidth]{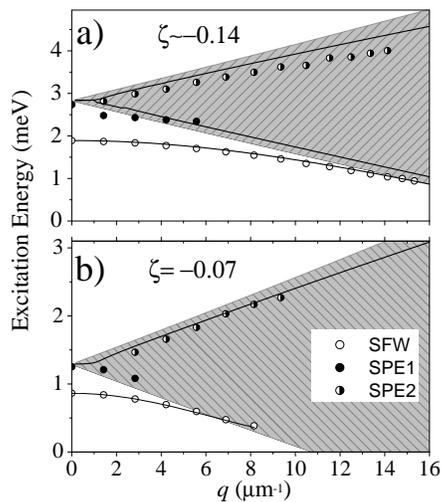}
\caption{Experimental dispersions of the SFW mode and features of
the SF-SPE continuum extracted from Fig.~\ref{fig:spectra}(a) and
(c). These, respectively, correspond to spin polarization values
$\protect\zeta=-0.14$ and $\protect\zeta=-0.07$. Experimental data
are compared with calculations: the dashed domain is the SF-SPE
continuum, edges of the domain are given for $T=0$ K. The lines
inside the domain (when available) are the peaks of the single
particle response [see Fig.~\ref{fig:SampB&Theory}(c)] calculated at
$T=2$ K. The line below the continuum is the OPM dispersion.}
\label{fig:disper}
\end{figure}

We now plot in Fig.~\ref{fig:disper} the dispersions of the SFW and
features of the SF-SPE extracted from Fig.~\ref {fig:spectra} and
compare with calculations. The spin polarizations in
Fig.\ref{fig:disper}(a) and \ref {fig:disper}(b) are respectively
$\zeta =-0.14$ and $\zeta =-0.07$. In the spectra of
Fig.~\ref{fig:spectra} and the dispersions in Fig.~\ref{fig:disper},
our data show clearly that the SFW mode propagates below the SF-SPE
continuum with a negative slope dispersion at energies that are in exact
agreement with the calculated OPM energy. Evidence is given that this
experimental mode is the OPM (or stiff mode of
Ref.\cite{FrustagliaPRB04}). Consequently, our observations are in
contradiction to this theory which predicted that the dynamical
\textit{s-d} exchange interaction between the Mn ions and the
electrons shifts the OPM above the SF-SPE continuum with a
positive dispersion. To resolve this contradiction, we have
introduced the Coulomb interaction between carriers to calculate the
coupled electron-Mn response. Details of the calculations are given in
Ref.\cite{PerezPRB10}.


The mixed response function is calculated adiabatically and leads to
the following electron spin susceptibility:

\begin{equation}
\left\langle \left\langle \hat{S}_{+,\mathbf{q}};\hat{S}_{-,-\mathbf{q}%
}\right\rangle \right\rangle _{\omega }=\frac{\left( \hbar \omega
-Z_{Mn}\right) \chi _{+}\left( \mathbf{q,\omega }\right) }{\hbar
\omega -Z_{Mn}-K\Delta \frac{2\pi \hbar ^{2}}{m^{\ast }Z^{\ast
}}\chi _{+}\left( \mathbf{q,\omega }\right) }
\label{eqCoulomb}
\end{equation}

\begin{figure}[tbp]
\begin{center}
\includegraphics[width=\linewidth]{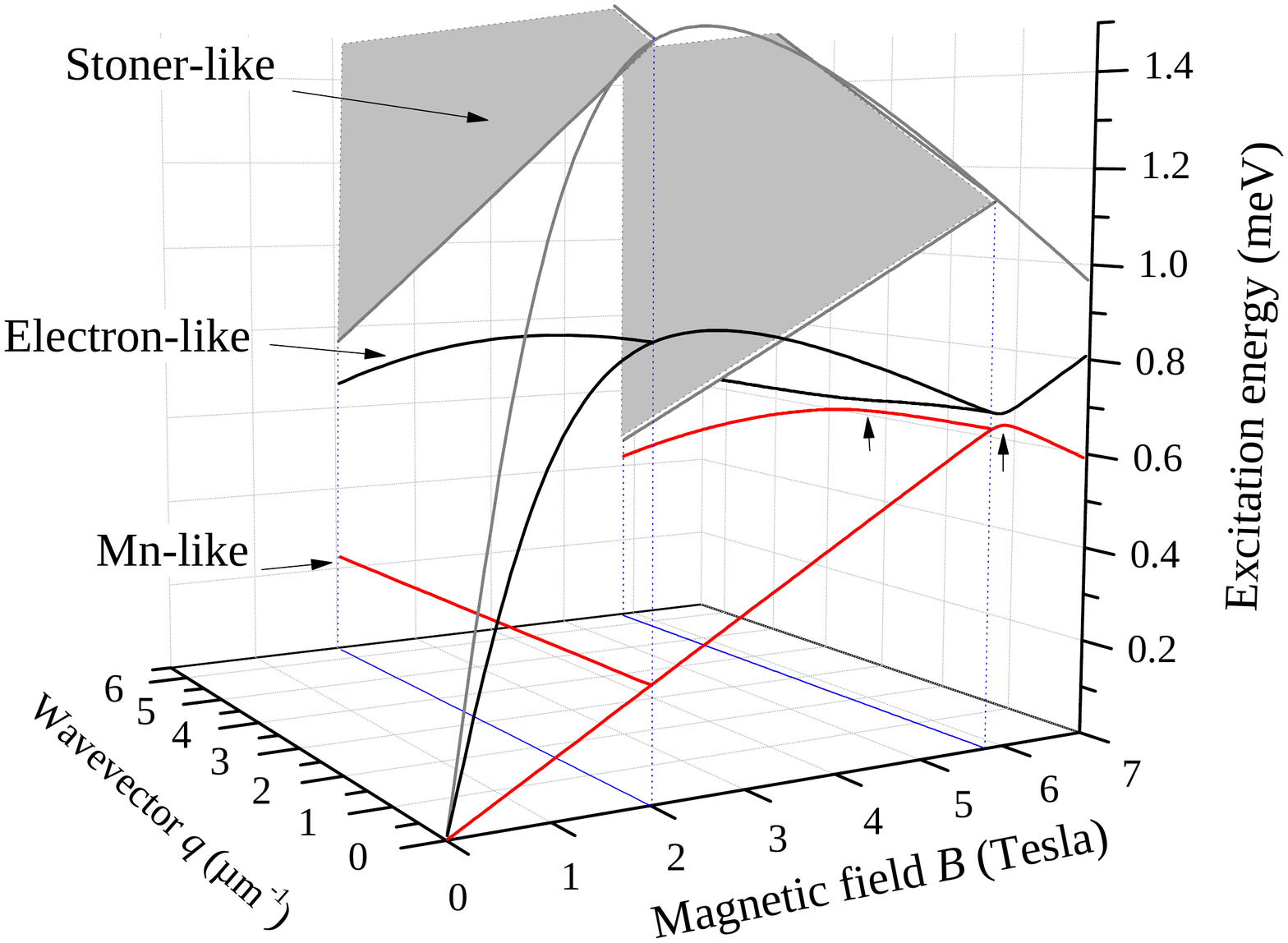}
\end{center}
\caption{(Color online) Electron-Mn modes as a function of both the
magnetic field $B$ and wavevector $q$. In the plane $q=0$, the two
lower curves are
the solutions $\hbar \protect\omega _{q=0}^{\pm }$ of Eq.(\ref
{ZonecenterModes}). The OPM is the higher energy solution. The
upper curve is the SF-SPE (Stoner-like excitations) degenerate to
$Z^{\ast }$. Out of the resonant field ($B_{R}=5.9T$) where the
modes anti-cross, the branches have electron or Mn character. At the
resonant field (indicated by a vertical arrow), the avoided gap is
about $25~ {\mu}$eV. In addition, dispersions $\hbar \protect\omega
_{q}^{\pm }$ obtained from Eq.(\protect\ref{eqCoulomb}) have been
plotted for $B=2$ T and $B=5.5$ T. For $B=2$ T the electron or Mn
characters of the mode do not change. However, for $B=5.5$ T the
modes anticross and an avoided gap appears at a finite $q_{R}$
(indicated by the vertical arrow). Modes exchange their slope dispersions and characters when $q$ varies accross $q_{R}$. Calculations
were done for $x_{\text{Mn}}=0.24\%,T=2~\text{K},n_{\text{2D}}=3.13\times
10^{11}\text{cm}^{-2}$ and a square quantum well of width
$w=15$ nm.} \label{FigAntiCross}
\end{figure}
where $\left\langle \left\langle \hat{S}_{+,\mathbf{q}};\hat{S}_{-,-\mathbf{q%
}}\right\rangle \right\rangle _{\omega }$is the linear response of
the transverse electron spin fluctuation observable
$\hat{S}_{+,\mathbf{q}}$ to the perturbation
$g_{e}{\mu}_{\text{B}}b_{+,\mathbf{q\omega }}$ created by a transverse
magnetic field of temporal and spatial Fourier components, respectively, given by
$\omega $ and $\mathbf{q}.$ \cite{Mauger83, UltrafastGaMnAs09} Coupled electron-Mn modes appear as
poles of $\left\langle \left\langle
\hat{S}_{+,\mathbf{q}};\hat{S}_{-,-\mathbf{q}}\right\rangle
\right\rangle _{\omega }$. In Eq.(\ref{eqCoulomb})
$Z_{\text{Mn}}=g_{\text{Mn}}{\mu}_{B}B+K$ is the Mn spin precession energy in the 
presence of the 2DEG equilibrium spin polarization, $g_{\text{Mn}}$
is the manganese $g$ factor; $\mu _{\text{B}}$ is the Bohr magneton;
$m^{\ast }$ is the electron effective mass; $\chi _{+}(
\mathbf{q},\omega )$ is the SP2DEG spin susceptibility;
\cite{PerezPRB2009} $\Delta =\alpha \gamma
x_{\text{Mn}}N_{0}\left\vert \left\langle \hat{S}
_{z}^{Mn}\right\rangle \right\vert \left( B,T\right) $ and
$K=\frac{1}{2}\frac{\alpha \gamma}{w}n_{\text{2D}}\left\vert \zeta \right\vert$
are the Overhauser and Knight shifts, respectively. $N_{0}\alpha
=220$ meV \cite{Gaj} is the \textit{s-d} exchange integral and
$\left\langle \hat{S}_{z}^{Mn}\right\rangle\left( B,T\right)$ is the
thermal average spin of a single Mn atom. $\gamma$ is the
probability of  finding an electron in the quantum well. Typical
orders of magnitude in our samples of $\Delta $ and $K$ are $\sim
10^{-3}$ and $\sim 10^{-6}$ eV, respectively. Interpretation of the
poles is as follows. In the frame of the Mn ions, the Mn precession
frequency is determined by the Zeeman energy of the Mn
($g_{\text{Mn}}\mu _{\text{B}}B$), the static mean-field
\textit{s-d} contribution from electrons ($K$), and the third term
is the "dynamical" Knight shift that originates from the electron
precession itself responding (through the $\chi
_{+}(\mathbf{q},\omega )$ response) to the Mn magnetic field
(proportional to $\Delta $). At $q=0$, $\chi
_{+}(\mathbf{q}=0,\omega)=-n_{\text{2D}}\zeta /\left( \hbar \omega
-Z_{e}\right)$, collective modes of Eq.(\ref{eqCoulomb}) develop
two modes: 
\begin{equation}
\hbar \omega _{q=0}^{\pm }=\frac{1}{2}\left( Z_{Mn}+Z_{e}\pm \sqrt{
(Z_{Mn}-Z_{e})^{2}+4K\Delta }\right), \label{ZonecenterModes}
\end{equation}
where $Z_{Mn}=g_{\text{Mn}}\mu _{\text{B}}B+K$. They open an avoided
gap at a specific magnetic field $B_{R}$ when $Z_{Mn}-Z_{e}=0$, as
illustrated in Fig.\ref{FigAntiCross} where $B_{R}=5.9$ T for
$x_{\text{Mn}}=0.24$.  The upper (resp. lower) energy branch is the OPM
(resp. in phase). For $B<B_{R}$, the OPM (resp. in phase) has an
"Electron like" character (resp. "Mn-like"), because it is mainly
dominated by the electron (resp. Mn) spin precession. In
resonance ($B=B_{R}$) modes have a complete mixed character. These
zone center modes were observed in Ref.\cite{TeranPRL2003} and
Ref.\cite{Vladimirova(2008)}. The resonance field depends mainly on
the Mn concentration and for the experimental situation of
Fig.\ref{fig:disper}, it is about 10.7 T. Figure \ref{fig:disper}
compares the dispersion of the OPM $\hbar \omega _{q}^{+}$,
calculated using Eq.(\ref{eqCoulomb}) and the parameters of sample A,
with the experimental one. Undoubtedly, the agreement with the data
is excellent without any fitting procedure. The dispersion of the
SFW mode $\hbar \omega_{q}^{SFW}$ defined as the unique pole of
$\chi _{+}(\mathbf{q},\omega )$ is also plotted. It appears that the
introduction of the \textit{s-d} dynamical coupling shifts $\hbar \omega _{q}^{+}$
negligibly from the pure electron mode
$\hbar \omega _{q}^{SFW}$ for that situation. However, the inclusion of the Coulomb interaction in Eq.(\ref{eqCoulomb}) 
positions the OPM below the SF-SPE continuum with a negative dispersion, contrary to what was found in 
Ref.\cite{FrustagliaPRB04}.

\begin{figure}[tbp]
\begin{center}
\includegraphics[width=\linewidth]{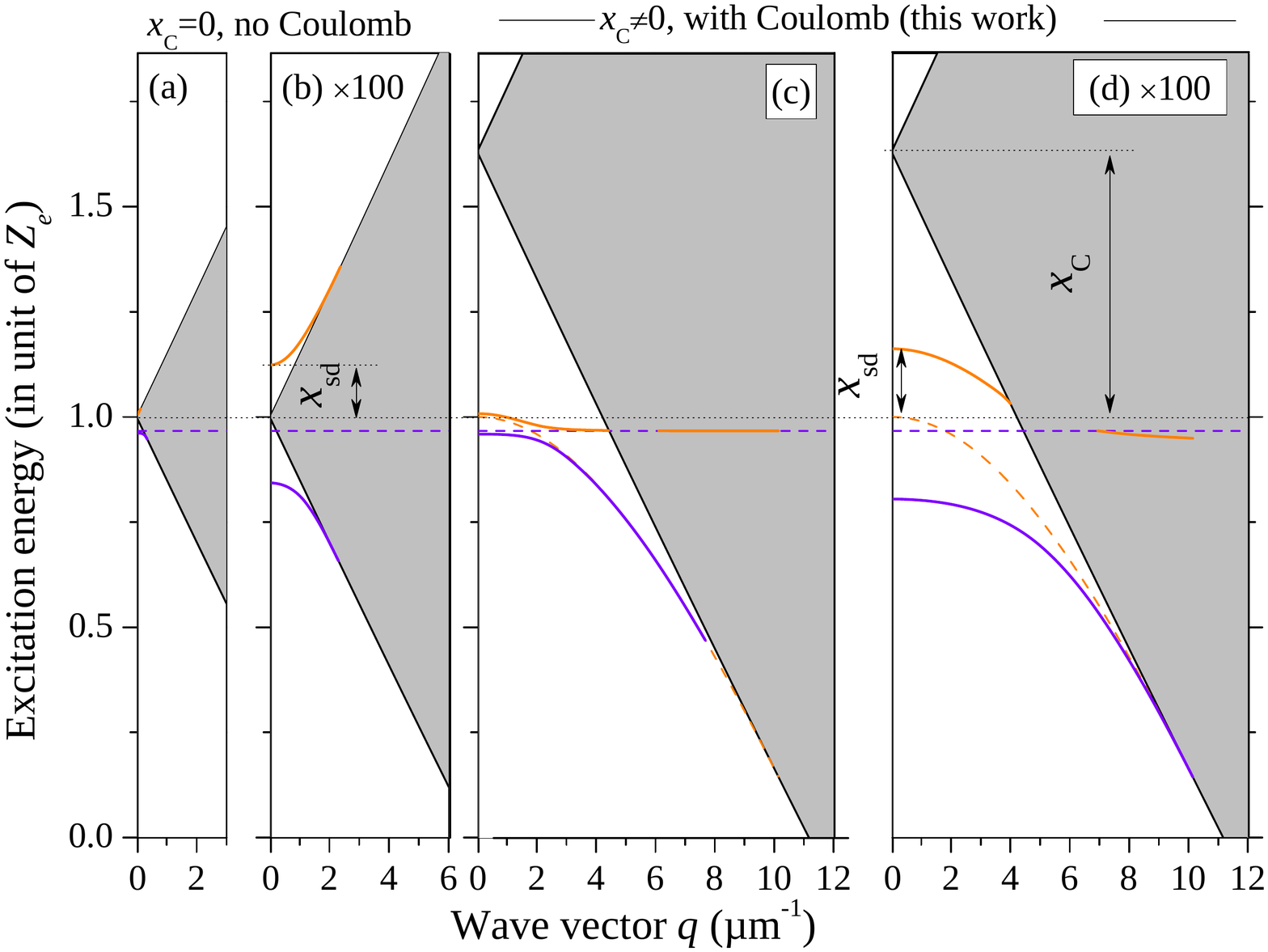}
\end{center}
\caption{(Color online) Illustration of the Coulomb and \textit{s-d}
dynamical shift on the electron-Mn modes dispersions. Dispersions
were calculated for $B=5.8$ T and the same sample parameters as in
Fig.\ref{FigAntiCross}. In (a) and (b), the Coulomb interaction
between electrons is suppressed, only the \textit{s-d} dynamical shift
applies as in Eq.(20) of Ref.~\cite{FrustagliaPRB04}. 
The dashed line is the uncoupled Mn mode energy ($Z_{\text{Mn}}$). In (c) and
(d), the Coulomb interaction between electrons is included as in
Eq.(\ref{eqCoulomb}) of this work. Dashed lines are the
uncoupled electron mode ($\hbar \omega _{q}^{SFW}$) and Mn mode
($Z_{Mn}$). At $q=0$ the SF-SPE energy is shifted to $Z^{\ast}$. In (b) and
(d) the \textit{s-d} dynamical coupling has been artificially magnified by a
factor of 100.} \label{Figxcxsd}
\end{figure}

Illustration of this phenomenon is provided in Fig.\ref{Figxcxsd}. In Fig.~
\ref{Figxcxsd}(a) and (b) the electron-Mn modes dispersion were
calculated using the denominator of Eq.(\ref{eqCoulomb}) by replacing the SP2DEG spin susceptibility with the non-interacting
equivalent, which is single particle response [see
Ref.\cite{PerezPRB2009}]. Such a calculation is equivalent to the
approach of Ref.\cite{FrustagliaPRB04}. Thus, for no Coulomb interaction between
electrons, the pure electron modes are the SF-SPE (poles
of the single-particle response) degenerate to $Z_{e}$ at $q=0$.
Inclusion of the \textit{s-d} dynamical coupling with the Mn introduces a
collective OPM propagating above the SF-SPE continuum (upper curve in Fig.
\ref{Figxcxsd}(a) and (b)). The relative "blue-shift" is labeled
$x_{sd}$. In presence of Coulomb interaction between carriers,
however, a collective electron mode (SFW) appears and propagates
below the SF-SPE continuum with a negative dispersion given by
$\hbar \omega_{q}^{SFW}$. The SF-SPE zone center energy is "blue
shifted" from $Z_{e}$ to $Z^{\ast}$ bye Coulomb-exchange. The
relative "red-shift" of the SFW mode with the SF-SPE is labeled
$x_{C}.$ Inclusion of the \textit{s-d} dynamical coupling with Mn introduces
a new shift which determines the final energy of the OPM.

Thus, we have separated two effects. First, Coulomb introduces a
shift between the collective electron mode and the SF-SPE energies
with a relative quantity quantifiable by $x_{C}=\left( Z^{\ast
}-Z_{e}\right) /Z_{e}$. Second, the \textit{s-d} dynamical coupling with the Mn
shifts this mode to $\hbar\omega _{q}^{+}$ with a relative shift
quantifiable as $x_{sd}=\left( \hbar\omega _{q=0}^{+}-Z_{e}\right)
/Z_{e}.$ Clearly, in the situation of Fig.\ref{fig:disper}:
$x_{C}\gg x_{sd}$.

We may ask if conditions can be found such that $x_{sd}>x_{C}$? If so, the OPM would propagate above the SF-SPE
continuum with a positive slope dispersion. We should distinguish two situations. For $B<B_{R}:$

\begin{equation}
x_{sd}\simeq \frac{K\Delta }{Z_{e}^{2}}\approx \frac{K}{Z_{e}}=\frac{Z^{\ast}}{Z_{e}}\frac{\alpha \gamma m^{\ast }}{4\pi \hbar ^{2}}
\frac{1}{w}  \label{xsd}.
\end{equation}
In resonance, however, $\left( B=B_{R}\right)$,

\begin{equation*}
x_{sd}\left( r\right) =\frac{\sqrt{K\Delta }}{Z_{e}}\simeq \sqrt{x_{sd}}. \label{xsd2}
\end{equation*}

\begin{figure}[tbp]
\includegraphics[width=\linewidth]{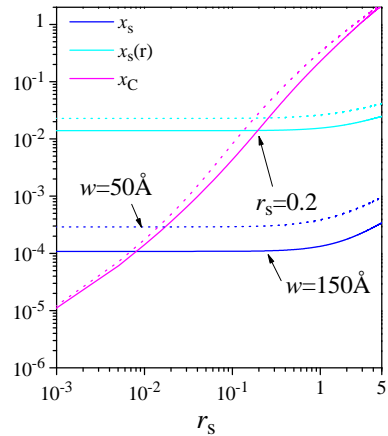}
\caption{(Color online) Comparison of the dynamical shift $x_{sd}$ calculated out or in resonance ($x_{sd}\left( r\right) $)
with the Coulomb shift $x_{C}$. Both are calculated as a function of the electron density for a fixed quantum well width, $w=50~\mathring{A}$ (dotted lines) and $w=150~%
\mathring{A}$ (full lines). The dynamical shift $x_{sd}\left( r\right)$ overcomes $x_{C}$ for a sufficiently low
electron density (respectively $r_{s}=0.2$ and $0.08)$. The out of resonance dynamical shift $x_{sd}$ overcomes the Coulomb
effect at much lower densities.}
\label{Shiftsfig4}
\end{figure}

In Figure~\ref{Shiftsfig4}, we compare $x_{sd}$ calculated out of
resonance and $x_{sd}\left( r\right)$ calculated in resonance with
$x_{C}$ for quantum well widths, $w=50~\mathring{A}$ and
$w=150~\mathring{A}$. Both $x_{sd}$ and $x_{sd}\left(r\right) $ are
calculated as a function of the electron density, plotted in terms
of $r_{s}$. As seen in Fig.~\ref{Shiftsfig4}, the
dynamical shift $x_{sd}\left(r\right)$ overcomes the Coulomb effect
at a sufficiently low electron density corresponding to $r_{s}=0.2$
and $0.08$. The out of resonance dynamical shift $x_{sd}$ overcomes
the Coulomb effect at much lower densities.

\begin{figure}
\begin{center}
\includegraphics[width=\linewidth]{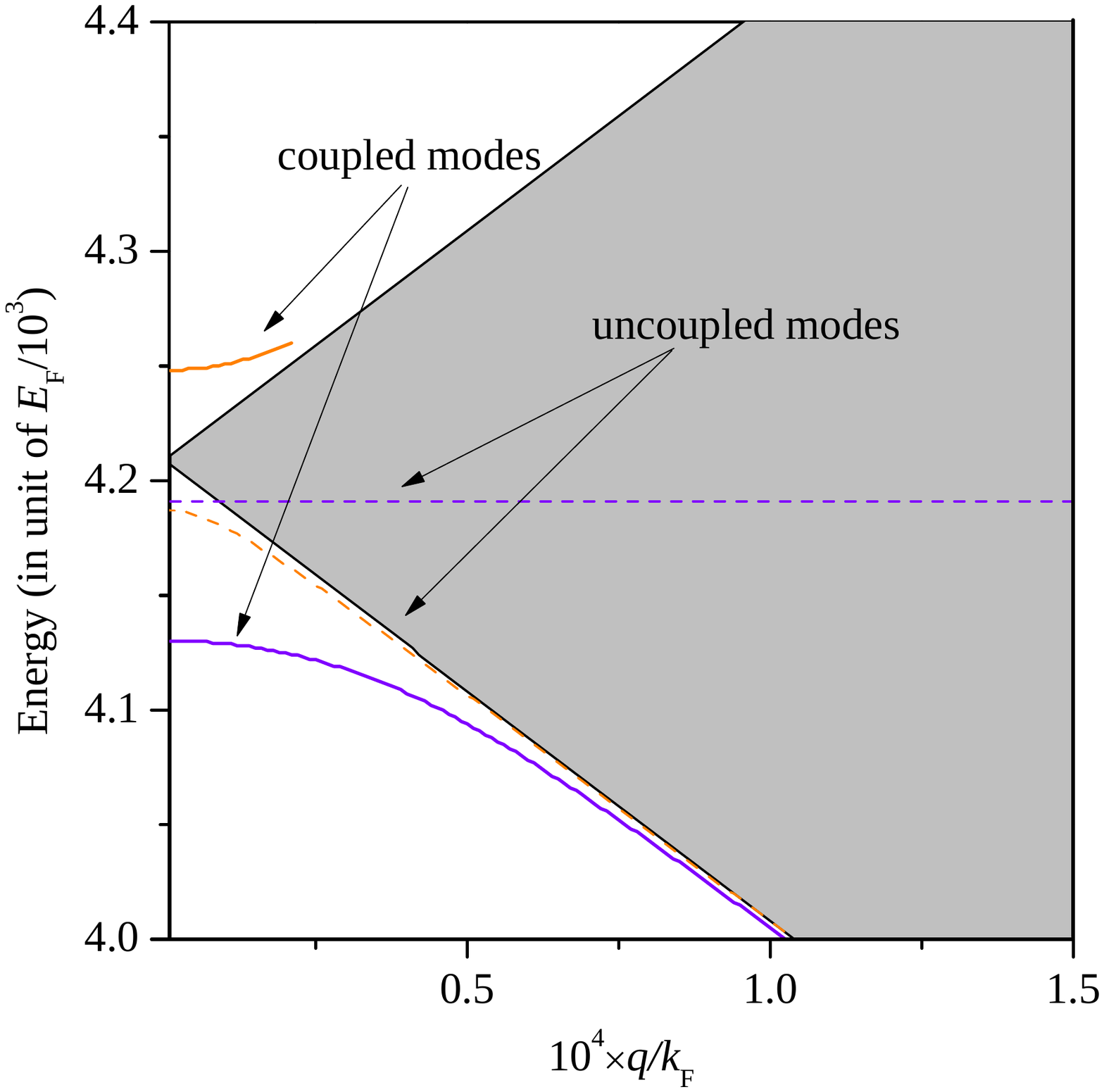}
\end{center}
\caption{(Color online) Dispersions of mixed spin modes calculated using the parameters of Fig.\protect\ref{Shiftsfig4}(a)
for $r_{s}=0.1$. The electron modes appears above the SF-SPE continuum. To get the effect, the Mn concentration is raised to
$x_{Mn}=10\%$ in order to get $\protect\zeta=0.2\%$. The external magnetic field is tuned to $B_{R}=130$ T. Here $w=150~%
\mathring{A}$ and $T=2$ K.}
\label{lowdensitydispersionsfig5}
\end{figure}

We now simulate a situation in which a prediction with $\hbar \omega
_{q}^{+}$ propagating above the SF-SPE is possible. We
have assumed the material parameters of Fig.~\ref{Shiftsfig4}, that is,
when the dynamical shift dominates the Coulomb shift. We show such
calculated dispersions of mixed spin modes in
Fig.~\ref{lowdensitydispersionsfig5} for $r_{s}=0.1$. Indeed, the
electron modes appear above the SF-SPE continuum. However, to obtain
such an effect, a significant spin polarization of $\zeta =0.2\%$
must be reached by raising the Mn ion concentration to
$x_{Mn}=10\%$. This means that an external magnetic field needs to
be tuned to a resonant field of $B_{R}=130$ T, an experimental
challenge! We note here that the small blue shift introduced to the
SFW energy by the dynamical coupling $x_{sd}$, explains the near
excellent agreement between our previous model and experiment. Since
the Coulomb effect dominates in our quantum well, the mixed modes
evolve essentially as an electron wave.

\section{Conclusions}

We have probed dispersions of spin flip waves and spin flip single
particle excitations in a spin polarized electron gas in a
Cd$_{1-x}$Mn$_x$Te quantum well using angle-resolved Raman
scattering. Our key result is that the SFW dispersion lies below the
SF-SPE continuum, contrary to predictions made by the theory of spin
waves in diluted magnetic systems. Analysis of our measurements with
a model accounting for the Coulomb interaction between carriers and
the dynamical response of Mn ions and the electron spin subsystem in
the spin wave propagator of the theory agrees well with our
experimental results. We have found that the Coulomb contribution
dominates over the dynamical response. We have investigated a regime
in which the dynamical coupling overcomes the Coulomb effect and
find this to occur at an extremely large external magnetic field and
for a high Mn ion concentration.

\begin{acknowledgments}
The authors would like to thank Jules Silembo, Michel Menant and
Silbe Majrab for machining parts of the magnetic cryostat and
electronics used in the experiments. Special thanks to Dimitri
Roditchev for experimental help. This work was partially supported
by the EPSRC and the CNRS. Additional financial support was provided
by the grant ANR GOSPININFO.
\end{acknowledgments}


\begin{thebibliography}{24}
\bibitem[{ang()}]{angle}
\bibinfo{note}{There is an additional factor of $\cos{\alpha/2}$ , where
  $\alpha$ is the angle between the scattered and incident light. The normal of
  incidence is taken as the zero. $\alpha$ is typically small ($\leq 2$
  degrees). Therefore $\cos{\alpha}\simeq 1$.}
\bibitem[{\citenamefont{Perez and et~al.}()}]{PerezPRB10}
\bibinfo{author}{\bibfnamefont{F.}~\bibnamefont{Perez}} \bibnamefont{and}
  \bibinfo{author}{\bibnamefont{et~al.}}, \bibinfo{journal}{To be published}.
\end{thebibliography}

\end{document}